\documentclass[twocolumn,english,aps,prl]{revtex4-1}
\usepackage[latin9]{inputenc}
\usepackage{amsmath}
\usepackage{amssymb}
\usepackage{graphicx}

\usepackage{babel}
\begin{document}

\title{Bose Gases Near Resonance:\\
renormalized interactions in a condensate}

\author{Fei Zhou and Mohammad S. Mashayekhi}

\affiliation{Department of Physics and Astronomy, University of British Columbia,
Vancouver V6T1Z1, Canada}

\date{12/02/11}
\begin{abstract}
Bose gases at large scattering lengths or beyond the usual dilute limit for long have been one of the most challenging problems
in many-body physics. In this article, we investigate the fundamental properties of a near-resonance Bose gas 
and illustrate that three-dimensional Bose gases become nearly fermionized near resonance when 
the chemical potential as a function of scattering lengths reaches a maximum and the atomic condensates lose
meta-stability. The maximum and accompanied instability are shown to be a precursor of the sign change of $g_2$, 
the renormalized two-body interaction between condensed atoms. That is $g_2$ changes from effectively repulsive to
attractive when approaching resonance from the molecular side, even though the scattering length is still positive. 
This occurs when dimers, under the influence of condensates, emerge at zero energy in the atomic gases at a finite positive scattering length.
We investigate the properties of near-resonance Bose gases via applying
a self-consistent renormalization group equation which is further subject to a thermodynamic boundary condition.
We also comment on the relation between the approach here and the diagrammatic calculation in an early article {\it [Phys. Rev. A {\bf 85}, 023620(2012)]}.
\end{abstract}
\maketitle

\section{introduction}

In the fascinating many-body world, the interplay between few- and many-body 
physics is known to play a very important role in a variety of systems and is instrumental to our understanding of
diversified emergent phenomena. The best known example perhaps is the superconductivity phenomenon 
where the Cooper bound states due to the Pauli blocking effect lead to pairing correlations and therefore superconductivity in
solids \cite{Bardeen57, Cooper56}.
Quantum degenerate gases of ultra cold atoms near Feshbach resonance that have been explored in many laboratories during the past decade
\cite{Chin10} are another fantastic platform to scrutinize
the intriguing interplay between few- and many-body physics.
Although far away from resonances, the physical properties of quantum gases can be 
well described by the existing dilute gas theories, very little, except in a few limiting cases, has been known about the fundamental properties 
of quantum gases near resonance beyond the usual dilute limit. 

The complexity of this topic to a large extent appears to be two-fold:

1) the role of few-body physics (two-body, three-body states etc) in many-body systems. In other words,
to what extent does the underlying few-body physics influence the many-body correlations and which channel dictates the many-body properties near resonance?

2) the effect of many-body background on the few-body structures. How are the few-body structures or multiple scatterings 
affected by the presence of many other identical particles?
 
These two issues can be treated successfully and separately in the usual dilute limit.
For instance,
in the leading order, one can neglect the effect of finite-density background atoms
on multiple scatterings or underlying few-body structures; the energy density can therefore be calculated perturbatively by assuming the few-body scatterings are given 
by their properties in the vacuum and applying the low density expansion. The effects of few-body structures on many-body physics can be
explored perturbatively. One can also further study the leading effect of quantum gases on dimers and trimers because the many-body states in the dilute limit are well-known.
However, near resonance, these two issues are generically entangled and ideally have to be addressed 
self-consistently, which usually becomes very challenging.

In this article, we make an attempt to understand the fundamental properties of Bose gases near Feshbach resonance
via examining the intriguing interplay between the few- and many-body physics in Bose gases at large positive scattering lengths.
For this purpose, we introduce a simple self-consistent renormalization-group-equation approach to address both sides of the coin.  
Many-body properties of a quantum gas are shown to influence the renormalization flow of few-body running 
coupling constants resulting in the change of the sign of the effective two-body interaction constants.
That in return completely dictates the many-body physics near resonance and leads to peculiar features in the chemical potentials.   
And we limit ourselves to the resonances with a very short effective range.

A key concept we are going to focus on is the effective interaction between condensed atoms near resonance.
The common believe is although the underlying short range resonance interaction has to be attractive,
at low energy scales the two-body interaction is effectively repuslive when the scattering length $a$ is positive.
The argument runs as the following.
The phase shift due to a short range attractive potential is given as
$\delta =-\arctan ka$
which yields $\delta=-ka$ when $k$ is small. The phase shift approaches $-\pi/2$ when $ka$ is much bigger than unity
but much smaller compared to $1/R^*$, $R^*(\ll a)$ is
the range of the attractive interaction.
The low energy phase shift $\delta=-ka$ turns out also to be the phase shift of any repulsive interaction with the same positive scattering length.
And so the atoms interact effectively repulsively if one is interested in the scatterings at small $k$.
However, the phase shifts of a short range attractive potential and a repuslive interation 
can become significantly different whenever $ka$ is bigger than unity;
for instance, for a hardcore potential with radius $R=a$, $\delta=-ka$ for all momenta, differing from the value of $-\pi/2$ for atttractive potentials when $ka \gg 1$.
Near resonance, $a$ approaches infinity and the issue of the effective
interaction between condensed atoms becomes very subtle. Indeed, near resonance we show that condensed atoms no longer interact  
with an effective repulsive interaction even when the scattering lengths are positive, a somewhat surprising conculsion to many.

The theory of dilute Bose gases has a long history.
The theory for weakly interacting bosons was established by Bogoliubov more than half century ago \cite{Bogoliubov47,Nozieres90}.
A properly regularized theory of dilute gases of bosons with contact interactions
was first put forward by Lee, Huang and Yang~\cite{Lee57} and later by Beliaev~\cite{Beliaev58} who
developed a field-theoretical approach.
Higher order corrections
were further examined in later years~\cite{Brueckner57,Wu59,Sawada59}. Since these
results were obtained by applying an expansion in terms of the small parameter $\sqrt{na^3}$ (here $n$ is the density and
$a$ is the scattering length),
it is not surprising that formally speaking each of the terms appearing
in the dilute gas theory diverges when the scattering lengths are extrapolated to
infinity. This aspect of the dilute gas theory,
to a large extent, is the main reason
why a qualitative understanding of Bose gases near resonance
had been missing for long.
In fact, the issue of quantum gases beyond the usual dilute limit has been one of the most outstanding unresolved problems in the quantum many-body physics.

A few preliminary experimental attempts have been recently made
to explore the properties of Bose gases beyond the dilute limit thanks to the availability of Feshbach resonances 
in many cold-atom laboratories~\cite{Navon11,Papp08,Pollack09,Inouye98}. 
It has been suggested that when approaching resonance from the molecular side,
Bose atoms appear to be thermalized within a reasonably short time,
well before the recombination processes set in,
so to form a quasi-static condensate.
Furthermore, the life time due to the recombination processes
is much longer than the many-body time scale set by
the degeneracy temperature which motivates us to explore the 
thermodynamic properties of the quasi equilibrium gases.
Here we intend to illustrate the intimate entanglement between few-body structures and many-body physics in a single approach. These illustrations suggest that most of
the surprising features of near-resonance Bose gases  
can be understood in terms of the spectrum flow of few-body structures such as dimers, trimers which can be strongly influenced by the interactions with the condensates.
The device we rely on to draw a qualitative picture is the renormalized interaction constant.

The two-fold complexities mentioned above are actually two sides of one coin and thus can not be completely separated, although
many theoretical attempts towards better understanding of near-resonance unitary gases have been focused on either of
these two aspects. For instance, the role of three-body physics in Bose gases in the dilute limit has been explored
and the contribution of Efimov states~\cite{Efimov70,Kraemer06,Zaccanti09,Pollack09a,Esry99} to the many-body energy density was estimated in the low density limit~\cite{Braaten02,Bulgac02}.
It was further discovered that the energy density of unitary Fermi gases can be directly related to
an asymptotic property of the reduced two-atom correlation function, which illustrates the important role of universal few-body physics 
in the many-body limit~\cite{Tan08,Zhang09,Braaten08,Werner08}. 
The interesting interplay between two-body physics and many-body physics explored in spectroscopy experiments was emphasized~\cite{Punk07,Baym07}.
A few separated efforts have also been made with an emphasis on the other side of the coin, i.e. how few-body structures respond to the presence of a quantum background, i.e. either statistical or dynamical effects.
These studies have led to the findings of anomalous dimers with tunable masses and new universal behaviours of Efimov states in quantum mixtures \cite{MacNeill11,Song11,Mathy11,Zinner11}. It remained to be seen
whether these predictions can be tested in future experiments. Furthermore, it was also shown that these anomalous few-body states can qualitatively modify the
low energy multiple scattering phase shifts~\cite{Song12}.

The approach outlined in this article is an alternative to the more elaborated diagrammatic resummation that was previously carried out 
by the authors and their collaborators in Ref.~\cite{Borzov11}.
The two approaches yield almost identical results.
In section II, we first carry out a simple scaling argument, as a caricature of resonating Bose gases, illustrating the relevance of fermionization in Bose gases near resonance 
and discuss the limitation of the coarse grain procedure. 
We also remark on a close relation between the effective two-body interactions and Lee-Huang-Yang corrections in Ref.~\cite{Lee57}.
In section III, we discuss the energetics of dimers and trimers in condensates and explore the implications on the many-body
physics. Especially, we point out that 
in addition to the fermionization phenomenon, an instability sets in at a positive critical scattering length as
a signature of formation of dimers in condensates.
We also show this aspect is a consequence of the sign-change of the renormalized two-body interactions between the condensed atoms;
the effect of condensates on the two-body interaction constant is investigated via taking into account the self energy of dimers and via imposing an infrared boundary condition for the renormalization flow.
In section V, we summarize the results of diagrammatic resummation.
We employ a set of self-consistent equations to address
the two issues listed at the beginning of the article. 
These self-consistent equations were first introduced by the authors and their collaborators in Ref.~\cite{Borzov11}.
To make it accessible to general readers, we leave the most technical part in the appendix and outline the framework of the approach and result in this section.   
In section VI, we conclude our studies.

\section{A caricature}

\subsection{Relevance of Fermionization: a scaling argument}

The energetics of Bose gases near resonance can be qualitatively understood via a coarse grain procedure which is more or less equivalent to the real space renormalization transformation. The simplest implementation of this is to first divide 
a quantum gas into N blocks each of which is of the size of
$\xi\times \xi \time \xi \times \xi$ where $\xi= 1/\sqrt{2m\mu}$ is the coherence length and $\mu$ is the chemical potential. 
Because the chemical potential is  a non-additive thermodynamic quantity, it is natural to define it as the change of energy when adding 
an additional atom to a particular block and the effect of other blocks is to set an appropriate boundary condition. 
Therefore, the chemical potential can be considered as the interaction energy between the added atom and existing atoms in the block.
If we further assume the interaction energy is dictated by a pairwise one, then

\begin{equation}
\mu=\epsilon_2(\xi,a; n) n \xi^3
\label{SC}
\end{equation}
where $\epsilon_2(L=\xi,a; n)$ is the characteristic interaction energy between two atoms in the block and $n\xi^3$ is the number of atoms in the block with $n$ being the number density.
This is a standard coarse grain procedure which relates a microscopic quantity $\epsilon_2(\xi, a;n)$ and a thermo-dynamic quantity, 
the chemical potential $\mu$.
The estimate of $\epsilon_2(\xi, a; n)$ itself is a full many-body problem that is usually very difficult to carry out.
In the dilute limit however, one can show that when two atoms interact in a box of size $\xi$, the probability of being scattered by the third particle
is negligible because the mean free path $l$ is proportional to $1/n 4\pi a^2$ ($a$ is the scattering length) much longer than $\xi$. In fact,

\begin{eqnarray}
\frac{\xi}{l} \sim \frac{a}{\xi}  \sim \sqrt{na^3}
\end{eqnarray}
which is small in the low density dilute limit.
So at least in this limit, we can approximate $\epsilon_2(L=\xi, a;n)$ 
as the energy of two interacting atoms $\epsilon_2(L=\xi,a; n=0)$ 
in an empty box of the size of the block.
If we assume this is also qualitatively correct even in the unitary limit, then we have a very simple self-consistent equation; the only knowledge we need to solve
this equation is how two atoms interact in a box of size $\xi$ at arbitrary scattering length $a$. $\epsilon_2(\xi,a;0)$ for a contact resonance interaction can be worked out, either by assuming
two atoms are in a harmonic trap of harmonic length $\xi$ or in a block of size $\xi$.  
The asymptotic behaviours are universal up to numerical prefactors. For two atoms in a block of size $L$,

\begin{equation}
\epsilon_2(L,a;0)=\begin{cases}
\frac{4\pi a}{ m L^3}(1+ C_1\frac{a}{L}+...) &  \mbox{when $a \ll L$}; \\
\frac{C_2}{2 m L^2} & \mbox{when $a \gg L$}.
\end{cases}
\label{2body}
\end{equation}
$C_{1,2}$ are two positive prefactors depending on the details of the block and are of little importance for our qualitative discussions here. 
It is important to notice that at resonance, $\epsilon_2(L=\xi, a=\infty;0)$ is finite and scales like the kinetic energy of an atom moving in an empty box of size $\xi$. 

Substituting the results in Eq. (\ref{2body}) into Eq. (\ref{SC}), one obtains the estimate of chemical potential. In the dilute limit, 

\begin{equation}
\mu=\frac{4\pi a n}{m} (1+C_1\sqrt{8\pi na^3}...);
\end{equation}
the correction to the first Hartree-Fock term is the leading finite size correction to the interaction energy and 
belongs to the well-known Lee-Huang-Yang effect. When $a$ approaches infinity on the other hand, this simple procedure leads to the prediction of fermionization.
That is 
\begin{equation}
\mu =\frac{1}{2m \xi^2}=\frac{C_2 n\xi}{2m}, \mbox{or} \mu \sim \frac{n^{2/3}}{2 m},
\end{equation}
which scales as the Fermi energy of a quantum gas with the same density and mass.
Although crude, the coarse grain shown here points to a phenomenon that was previously seen in a few numerical calculations. Given that it is very simple,
we consider it quite a success. The relevance of fermionization to Bose gases near resonance was observed in a few theoretical studies
\cite{Cowell02,Song09,Lee10,Diederix11}.

This aspect of Bose gases near resonance is also an essential feature of Tonks-Girardeau gases or hardcore bosons in one dimension \cite{Girardeau60,Lieb63}.
The one-dimensional Bose fluids were later further studied using the Luttinger liquid formulation \cite{Haldane81}; the renormalization of
microscopic parameters also plays a very important role in the construction of that theory although the implementation was carried out very differently from
what we are going to focus on below.

\subsection{Running two-body interaction constants}

But how good is the starting point that near resonance we can approximate $\epsilon_2(L, a;n)$ as the two-particle interaction in an empty box completely neglecting the effect of
many other identical particles?  To address this, we estimate $\epsilon_2(L,a;n)$, the interaction of two atoms in a box of size $L$ via employing
a more sophisticated approach, the real-space renormalization transformation (RSRT) which further takes into account
the many-body effect on $\epsilon_2(L,a; n)$. 
This approach indicates the fermionization
must not be the whole story. 

Consider, instead of $\epsilon_2(L,a;0)$ discussed above,

\begin{eqnarray}
g_2(L,a;n)= \epsilon_2(L,a;n) L^3
\end{eqnarray}
which is the effective strength of the short range two-body interaction.
Again we divide the length scales in RSRT into two regimes that are separated by
 $\xi$: the short distance regime in which two- and few-body physics dominates and the long wave length regime where the many-body collective effect dominates. 
$\xi$ defines the interface where the microscopic few-body parameter $g_2$ at shorter distance needs to match the macroscopic coarse grain condition.

So at scales smaller than $\xi$, we can employ the RSRT of the two-body running coupling constant
$g_2(L,a;0)$ in vacuum to monitor the effective interaction; the finite density has very little effect in this regime i.e. 
$g_2(L,a;n)=g_2(L,a;0) +O(L/\xi)$.
At larger distances, because $g_2$ defined here is subject to
a thermodynamic constraint of the chemical potential at $L \sim \xi$, 
the effect of condensate on $g_2$ or $\epsilon_2$ is to impose 
a boundary condition on the flow of $g_2(L, a; n)$ via the coarse-grain condition in Eq. (\ref{SC}). 
And in this approach, the collective physics at scales larger than $\xi$ 
influences the flow solely through a simple boundary condition. 

\begin{figure}
\includegraphics[width=\columnwidth]{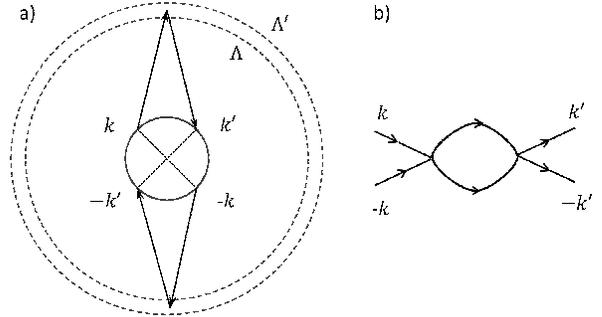}
\caption{(Color online) a) Schematic of the renormalization of the low energy on-shell scattering amplitude.
An initial state $({\bf k},-{\bf k})$ on the inner sphere is first scattered into off-shell high energy states $({\bf p},-{\bf p})$
represented by the shell region between two outer momentum spherical surfaces (dashed) before finally being scattered backed into 
$({\bf k}', -{\bf k}')$  on the inner sphere. Here
the outer dashed spherical surface is defined by $\Lambda'$ and the inner dashed one by $\Lambda$;
$\Lambda \leq  |{\bf p}| \leq \Lambda'$. 
b) The one-loop diagram that leads to the renormalization equation below.
The internal lines are for states within the shell region described in a).
Each vertex stands for the two-body interaction $g_2(\Lambda')$.
}
\label{fig1}
\end{figure}

Practically, since $L$ defines the size of micro-blocks in the renormalization procedure, it therefore defines the momentum cut-off via
$\Lambda=L^{-1}$. The transformation from $L$ to $L'$ is equivalent to rescale the momentum from $\Lambda$ to $\Lambda'=L'^{-1}$.
To obtain the running coupling constant, one can use the standard momentum-shell renormalization procedure to track the transformation
from the original $g_2$ to the new one $g'_2$ when the Hilbert space or the momentum cut-off is rescaled from $\Lambda'$ to 
$\Lambda=\Lambda' -\delta \Lambda$ (See Fig. \ref{fig1}).

The reduced two-body Hamiltonian we use for this illustration is

\begin{eqnarray}
H_{2-body}&=& H_< + H_> +H_{><}\nonumber \\
H_{<}&=&\sum_{\bf k} \epsilon_{\bf k} b^\dagger_{\bf k}b_{\bf k} +\frac{g_2}{2\Omega}\sum_{{\bf k},{\bf k}'}
(b^\dagger_{\bf k}b^\dagger_{-\bf k} 
b_{\bf k'}b_{-\bf k'} +h.c.) \nonumber \\
H_{>}&=&\sum_{\bf p} \epsilon_{\bf p} b^\dagger_{\bf p}b_{\bf p} +\frac{g_2}{2\Omega}\sum_{{\bf p},{\bf p}'}
(b^\dagger_{\bf p}b^\dagger_{-\bf p} 
b_{\bf p'}b_{-\bf p'} +h.c.) \nonumber \\
H_{><}&=& \frac{g_2}{2\Omega}\sum_{{\bf k},{\bf p}} 
b^\dagger_{\bf k}b^\dagger_{-\bf k} 
b_{\bf p}b_{-\bf p} +h.c.. 
\label{H2-body}
\end{eqnarray}
Here 
$b^\dagger_{\bf k}(b_{\bf k})$ is the creation (annihilation) operator for a Bose atom with momentum ${\bf k}$,
$\Omega$ is the volume.
The sum in $H_{<}$ is over momenta $|{\bf k}|,|{\bf k}'|$ smaller than  $\Lambda$ and
the sum in $H_>$ is over the states within the shell in Fig. \ref{fig1}, i.e. $ |{\bf p}|, |{\bf p}'| \in [\Lambda, \Lambda']$.
$H_{><}$ describes the off-shell scattering from low momentum states with $|{\bf k}|<\Lambda $ into the high energy states with momenta ${\bf p}$ within the shell and vice versa.
This interaction can induce an effective scattering between low momentum states $({\bf k}, -{\bf k})$ and $({\bf k}', -{\bf k'})$, 
 $|{\bf k}|, |{\bf k}'| \leq \Lambda$, via virtual states $({\bf p}, -{\bf p})$ within the shell.

When rescaling, the states within the shell thus lead to an additional contribution to the two-body interaction in $H_<$.
One can obtain the beta-function for the renormalization equation diagrammatically.
This calculation is very similar to the T-matrix calculation except one should restrict to the virtual states within
the shell between $\Lambda'$ and $\Lambda=\Lambda'-\delta \Lambda$.
The diagram in Fig. \ref{fig1} represents such an additional contribution to the effective two-body interaction,

\begin{eqnarray}
-i \delta g_2(\Lambda') &=&i^4  g^2(\Lambda') \int^{'} \frac{d{\bf p}}{(2\pi)^3}\int \frac{d\epsilon}{2\pi}G^0(\epsilon, {\bf p})G^0(-\epsilon, -{\bf p})\nonumber \\
G^0(\epsilon, {\bf p})& = &\frac{1}{\epsilon -\frac{{\bf p}^2}{2m}+i\delta}.
\label{2bodyRG}
\end{eqnarray}
Here the momentum integral $\int^{'}$ is over the states within the shell shown in Fig. \ref{fig1}, i.e. $\Lambda' > |{\bf p}|> \Lambda$.

After carrying out the energy and momentum integrals, 
one can easily find the transformation for the two-body interaction constant

\begin{eqnarray}
g_2(\Lambda' -\delta \Lambda)=g_2(\Lambda')-\frac{m}{2\pi^2} g^2_2(\Lambda') \delta \Lambda +O(\delta \Lambda^2).
\end{eqnarray}

The transformation of $g_2$ under the real space rescaling can be obtained by converting $\Lambda$ to $L^{-1}$.
For a quantum gas with a finite density, we
therefore have ($\tilde{g}_2=g_2/L$)

\begin{eqnarray}
&& \frac{\partial \tilde{g}_2(L)}{\partial \ln L^{-1}} = \frac{m}{2\pi^2} \tilde{g}_2^2(L)+\tilde{g}_2(L),\nonumber \\
&& 
\tilde{g}_2(L=R^*)=\frac{U_0}{R^*},  
\tilde{g}_2(L=\xi)= \frac{\mu}{n \xi}.
\label{RGmany}
\end{eqnarray}
The boundary condition at $L=\xi$ is exactly the condition in Eq. (\ref{SC}), i.e. at scale $\xi$ the microscopic running coupling constant has to match
the thermodynamic constraint suggested by $\mu$, assuming the main contribution to $\mu$ is from the two-body interaction $g_2(L,a;n)$.
At a very short distance $R^*$, the boundary condition is set by $U_0$, the strength of the {\em bare} two-body short range attractive interaction with range $R^*$.
And for the resonance phenomena we are interested in,
$R^*$ is always much smaller than $a$. 

By contrast, in a vacuum the coupling constant $g_2$ should flow to the value of $4\pi a/m$, the 
standard form of the two-body effective interaction, or 

\begin{eqnarray}
&& \frac{\partial \tilde{g}_2(L)}{\partial \ln L^{-1}} = \frac{m}{2\pi^2} \tilde{g}_2^2(L)+\tilde{g}_2(L),\nonumber \\
&&  \tilde{g}_2(L=R^*)=\frac{U_0}{R^*},  
\tilde{g}_2(L \rightarrow \infty,a; n=0) = \frac{4\pi a}{m L}.
\label{RGfew}
\end{eqnarray}
For a bare attractive two-body interaction with strength $U_0(<0)$ and range $R^*$, 
the boundary condition at $L=\infty$ in
Eq. (\ref{RGfew}) establishes a well-known relation between $U_0$ and the scattering lengths $a$.
$g_2=L \tilde{g}_2(L)$ as the solution to Eq.\ref{RGfew} can be expressed in terms of $a$, 

\begin{eqnarray}
{g}_2(L, a; n=0)= \frac{4\pi a}{m}\frac{1}{1 -\frac{2 a}{\pi L}}.
\label{fsg2}
\end{eqnarray}
Obviously, $g_2$ appears to be repulsive only in the limit of long wave length 
when $L \gg a$. At short distances $R^* < L \ll a$, 

\begin{eqnarray}
g_2(L) \rightarrow - \frac{2\pi^2 L}{m}
\label{univer}
\end{eqnarray}
is negative and independent of $R^*$ or $a$. Eq. (\ref{univer}) indicates a universal form of the two-body running coupling constant that induces resonance scatterings.
This crossover from repulsive to attractive interactions happens at $L^* \sim a$. 

One can further show that for a repulsive interaction that leads to the same
zero energy scattering length $a$, $g_2$ also flows toward the value of $4\pi a/m$ when $L$ is much longer than the range
of interaction. For instance for a hardcore potential with $a=R^*$ where $R^*$ is the radius of hardcore, one obtains the same expression as Eq. (\ref{univer}) except
that the range of $L$ is $ 2 a/\pi < L < \infty$; and not surprisingly, $g_2$ in this case is repulsive for arbitrary length scales. 

So only in the long wave length limit, the attractive interaction with positive scattering lengths 
yields the same physics as the repulsive ones even though at short distances they are distinctly different.
At the zero energy, the effective interaction is $4\pi a/m$, repulsive as far as $a$ is positive disregarding
whether the bare interactions are repulsive or attractive.
This for long has been a common belief in the field of cold atom physics.

As we will see below, this no longer holds near resonance when the many-body renormalization effects due to condensed atoms are further taken into account.
The reason for this is the low energy window where we can approximate the resonance interaction as a repulsive one (which is of order $1/ma^2$) gets so narrow that the effect of condensates
on the two-body coupling becomes particularly pronounced near resonance.

\begin{figure}
\includegraphics[width=\columnwidth]{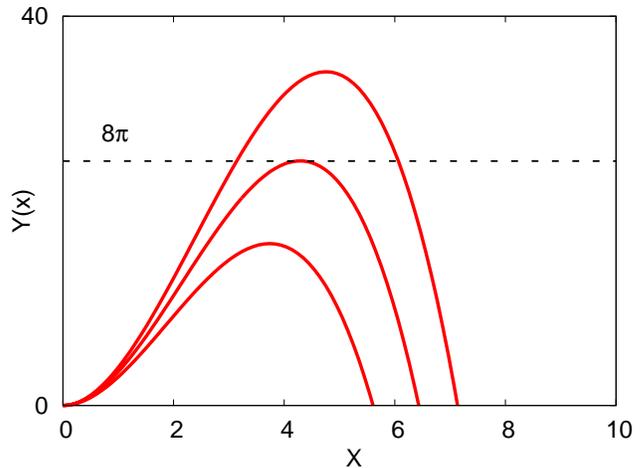}
\caption{(Color online) The solution to the self-consistent boundary condition, Eq.(\ref{SCE}).
The solution is obtained by solving $Y(x)=x^2/(n^{1/3}a)- (2/\pi) x^3=8\pi$,
where $x^2=2m\mu/n^{2/3}$. 
From the top to bottom are $Y(x)$ for $a < a_{cr}$, $a=a_{cr}$ and $a > a_{cr}$.
At $a_{cr}$, the equation has only one solution and above $a_{cr}$ there are no real solutions.
}
\label{fig2}
\end{figure}

The renormalization-equation approach had been previously applied to analyze the effective field theories for few-body scattering phenomena \cite{Kaplan98}.
They were also employed to
identify the coupling constants and quantum phases in the field theory models for
the lower branch unitary gases with a lot of successes \cite{Nishida06,Veillette07,Nikolic07}.
It was later employed to explore the physics of geometric resonances and confinement induced scattering phenomena \cite{Cui10}.
Our application to Bose gases is perhaps another excellent example to demonstrate
that the simple and generic approach of renormalization can lead to some surprising breakthroughs.

Eq. (\ref{RGmany}) is a RSRT equation which satisfies Eq. (\ref{SC}) and yields an estimate of $\epsilon_2(\xi, a;n)$ or $g_2$.
The boundary condition leads to a self-consistent equation for $\mu$.
When expressing in terms of $a$ using the solution to Eq. (\ref{RGfew}), one finds

\begin{eqnarray}
\mu = n \frac{4\pi a}{m}\frac{1}{1-\frac{2}{\pi} \sqrt{2m \mu} a}.
\label{SCE}
\end{eqnarray} 
Again in the limit where $a$ is much less than $\xi$, the equation yields the Hartree-Fock energy plus the correction of Lee-Huang-Yang character;

\begin{equation}
\mu=\frac{4\pi a n}{m} (1+ \frac{4\sqrt{2}}{\sqrt{\pi}}\sqrt{na^3}+...).
\end{equation}

Another solution of $\mu$ scales as $1/a^2$ consistent with the binding energy of lower branch molecules and we don't consider here.
In the unitary limit however, the equation not only indicates fermionization but also suggests a critical point beyond which 
there are no real solutions to the equation. This is
most obvious when $a$ is infinity and $g_2$ becomes {\em negative}. 
This property of Eq. (\ref{SCE}) is
illustrated in Fig. 2.
One can show that at the critical point,

\begin{eqnarray}
n^{1/3}a_{cr}=\frac{1}{6}\pi^{1/3}, \mu_{cr}=2 \pi^{4/3} \frac{n^{2/3}}{m};
\end{eqnarray}
and two real solutions merge into a single one. Beyond this point, 
the equation yields to a complex solution to the chemical potential.

The RSRT suggests an important feature that is absent in the simplest coarse grain approach (Eq. (\ref{SC}),(\ref{2body})). 
It turns out that near resonance, there is a substantial modification of
the underlying two-body physics, i.e. dimer energetics and therefore the interaction energy between condensed atoms $\epsilon_2$;
it can no longer be  justified to approximate $\epsilon_2(\xi,a; n)$ as the interaction energy in an empty box or at zero density. 
In fact as shown below, the uplifted dimers (towards condensates) cause an instability of atomic condensates when approaching resonance 
from the molecule side.
The emergency of the imaginary part of the chemical potential beyond the critical point signifies a hybridization between atoms and molecules which is missing in
the simplest coarse grain argument. In the next section, we further elaborate on this fascinating aspect of Bose gases near resonance.

\section{Dimers and trimers in a condensate:\\ 
the spectrum flow}

How are the dimers or trimers formed in the presence of a condensate or of a quantum gas?
In the context of quantum mixtures, there have been a few attempts to answer this question: how are the few-body structures affected by the presence of a Fermi surface \cite{MacNeill11,Song11,Mathy11,Zinner11}?
Surprisingly so far little efforts have been made to understand the dimers and trimers in the presence of a condensate, partially because the background of a condensate is more dynamical compared to that of
a Fermi sea. Since this plays a critical role in the interplay between few- and many-body physics that interests us, here we make an effort to estimate the effect.

It is possible to solve the two-body and three-body S-matrices in the presence of the many-body effect due to the self-energy.
Assuming the self-energy of quasi-particles is $\Sigma$, one finds that for two incoming atoms with momentum ${\bf p}, -{\bf p}$
scattered into ${\bf p}',{-\bf p}'$ and with total frequency $E$, 

\begin{widetext}

\begin{eqnarray}
G_2(E; {\bf p}, {\bf p}')=U_0 + U_0 \int \frac{d^3{\bf q}}{(2\pi)^3} \frac{1}{E-2\eta-2\epsilon_{\bf q}+i\delta^+}G_2(E;{\bf q}, {\bf p}')
\end{eqnarray}
\end{widetext}
where $\eta=\Sigma-\mu$.
A diagrammatic representation is given in Fig. 3(a).
In the dilute limit, $\eta=\mu=4\pi a n/m$; in general, $\Sigma$, $\mu$ and $\eta$ are unknown and need to be determined self-consistently later on.
For now, we simply assume that $\eta$ is a given parameter. 

$G_2(E;{\bf p},{\bf p}')=G_2(E)$ as a result of the short range interaction and
note that when $\eta=0$ or in vacuum, 
$G^0_2(E=0)=4\pi a/m$ (superscript $0$ indicates the case of vacuum) and $G^0_2(E)=4\pi a/m (1-ia\sqrt{mE})$.
One can then show that in this approximation,

\begin{equation}
G_2(E)=G^0_2(E-2\eta);
\end{equation}
the pole of $G_2(E)$ is shifted from the pole in vacuum by $2\eta$.
The pole defines the dimer binding energy and so

\begin{equation}
\epsilon_{D}=\epsilon^0_{D}+2\eta
\end{equation}
where $\epsilon_{D}$ and $\epsilon^0_D$ are the binding energy of dimers in the presence of a condensate and in vacuum respectively.
At a given positive scattering length, the dimer spectrum flows (in the energy space) towards the zero energy where the condensate lives as one increases the $\eta$.

One can also calculate the amplitude of three-body scatterings corresponding to the processes described in Fig. 3(b).
We consider a general case where three incoming momenta are
${\bf k}_1={\bf p}/2-{\bf q}$, ${\bf k}_2={\bf p}/2+{\bf q}$ and
${\bf k}_3=-{\bf p}$, and outgoing ones are
${\bf k}'_1={\bf p}'/2-{\bf q}'$, ${\bf k}'_2={\bf p}'/2+{\bf q}'$ and
${\bf k}'_3=-{\bf p}'$.
The scattering amplitude between theses states can then given by $A_3(E;{\bf p}, {\bf p}')$ where ${\bf q}$ and ${\bf q'}$ doesn't enter explicitly;
it represents the sum of diagrams identical to Fig. 3(b).

It is more convenient to work with the reduced amplitude
$G_3(E;{\bf p})=A_3(E;{\bf p},0)$ where ${\bf p}'$ is already taken to be zero. 
$G_3(E;{\bf p})$ itself obeys a simple integral equation as can be seen by listing the terms in the summation explicitly.
The diagrams in Fig. 3(b) yield (see Appendix; the mass is set to be one, i.e. $m=1$),

\begin{widetext}
\begin{eqnarray}
\frac{1}{4} G_{3}(E,p) & = & - \frac{1}{2}K(E-3\eta;p,0) +
 \frac{2}{\pi}\int dq\frac{K(E-3\eta;p,q)q^{2}}{\sqrt{\frac{3}{4}q^{2}+3\eta-E+i\delta^+}-\frac{1}{a}}\frac{-1}{q^{2}+3\eta-E}\\
& + & \left(\frac{2}{\pi}\right)^{2}\int dqdq'\frac{K(E-3\eta;p,q)q^{2}}{\sqrt{\frac{3}{4}q^{2}+3\eta-E}-\frac{1}{a}}\frac{K(E-3\eta;q,q')q'^{2}}{\sqrt{\frac{3}{4}q'^{2}+3\eta-E}-\frac{1}{a}}
\frac{-1}{q'^{2}+3\eta-E}+\cdots
\end{eqnarray}
\end{widetext}
where $K(E-3\eta; p,q)$ is the kernel defined as

\begin{eqnarray}
K(E-3\eta;p,q)=\frac{1}{pq}\ln \frac{p^2+q^2+pq+3\eta-E}{p^2+q^2-pq+3\eta-E},
\end{eqnarray}
The sum of the above infinite series leads to the following integral equation
of $G_{3}$ as 
\begin{widetext}
\begin{equation}
G_{3}(E,p)= - 2 K(E-3\eta;p,0)
+\frac{2}{\pi}\int dq\frac{K(E-3\eta;p,q)q^{2}}{\sqrt{\frac{3}{4}q^{2}+3\eta-E}-\frac{1}{a}}G_{3}(E,q).\label{G3IE}
\end{equation}
\end{widetext}
When $\eta=0$ as in vacuum, this equation is identical to an integral equation previously obtained in an atom-dimer model to study the renormalized three-body forces \cite{Bedaque99}.   
Comparing to $G_3(E,p)$ in vacuum when $\eta=0$, again one finds that the energy of a trimer in a condensate $\epsilon_T$ is related to $\epsilon^0_T$, its vacuum value via

\begin{eqnarray}
\epsilon_T=\epsilon^0_T+3\eta.
\label{trimer}
\end{eqnarray}
What Eq. (\ref{trimer}) shows is a simple fact of a condensate. If all the finite momentum atoms have a mean-field energy shift $\Sigma -\mu$ with respect to condensed atoms,
the energy of few-body bound states (with finite $k$ components) experiences the corresponding energy shifts.
As a consequence, when $\epsilon_T=0$, we should expect that the three-body forces in a condensate should be divergent.
This was observed numerically in Ref. \cite{Borzov11}; the three-body potential is divergent when $3\eta =-\epsilon_n$ where $\epsilon_n$ are the Efimov eigen values with
$n=1,2,3,...$.

What is the consequences of the spectrum flow or the energy shift due to the condensate? 
The main consequence is that
in a condensate, a dimer crosses the zero energy, or the energy of condensed atoms at a positive critical scattering length $a_{D}$ or $\epsilon_D=0$ 
when

\begin{equation}
2\eta(a_D) =\frac{1}{ma^2_{D}},
\label{AD}
\end{equation}
where $\eta$ itself is a function of $a_D$.
By contrast, in vacuum, a dimer crosses the zero energy or the scattering threshold at resonance or $a=\infty$.
If we simply apply the Hartree-Fock approximation $\Sigma=2\mu=8\pi a n/m$, we find
$\eta=\mu$ and 

\begin{equation}
n^{1/3} a_D=(1/8\pi)^{1/3}.
\end{equation}
Beyond this point, one has to take into account the hybridization between atoms and molecules.
The dimer formation in condensates was previously studied in a random-phase approximation; those results are qualitatively consistent with the picture painted here \cite{Yin08}.
Pairing instability and formation of molecules in the upper branch Fermi gas was emphasized in Ref. \cite{Pekker11}.

Since it is necessary to have molecules below condensates so that condensed atoms have effective repulsive interactions, the penetration of dimers into the condensate
implies a change of the sign of interactions, from repulsive to attractive ones that can lead to a potential instability.  Below we further amplify this aspect.

\begin{figure}
\includegraphics[width=\columnwidth]{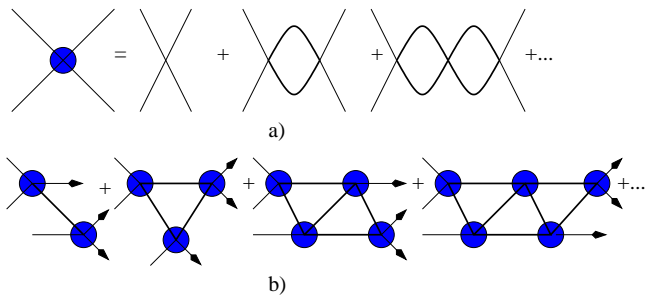}
\caption{(Color online) Diagrams for the calculations of dimer and trimer energy in a condensate.
a) is for two-atom channels. Each solid internal line here is for a Green's function of non-condensed atoms with the self-energy effect taken into account,
$G^{-1}(\epsilon, {\bf k})=\epsilon-\epsilon_{\bf k} -\Sigma+\mu+i\delta$.
b) is the loop diagrams for the three-atom channel. 
}
\label{fig3}
\end{figure}

\section{Sign change of $g_2$:\\ a consequence of spectrum flow}

In a condensate,
the low energy two-body interaction constant is renormalized not only by the virtual scattering states as in vacuum but also by the interactions with the condensed atoms.
The later effect is many-body in nature. Below we focus on the particular many-body effect related to the self-energy of
virtual states and include this in the renormalization procedure.
The self-energy of non-condensed atoms is due to scatterings by condensates and depends on the interaction strength and density of atoms.

We now apply a self-consistent renormalization group equation (RGE) to investigate this issue.
To study the coupling constant, 
we start with an assumption that the self-energy and the chemical potential of non-condensed particles 
are already given as $\Sigma$ and $\mu$.
The simplest Hartree-Fock Green's function for virtual atoms is of the form 

\begin{eqnarray}
G(\epsilon,{\bf p})=\frac{1}{\epsilon-\frac{{\bf p}^2}{2m} -\Sigma+\mu+i\delta}.
\end{eqnarray}
We then calculate $g^0_2(\Sigma,\mu)$, the zero energy effective interaction between condensed particles for a given $\eta=\Sigma-\mu$ 
using a very similar procedure as that in section II except $G^0$ in Eq.(\ref{2bodyRG}) now should be replaced with $G$ defined here.
The corresponding renormalization group equation (RGE) for the running couple constant $\tilde{g}_2=g_2 (\Lambda) \Lambda_h$ can be found to be,

\begin{eqnarray}
\frac{\partial \tilde{g}_2(\Lambda_h)}{\partial \ln \Lambda_{h}} &=& \frac{m}{2\pi^2} \tilde{g}_2^2(\Lambda_h)+\tilde{g}_2(\Lambda_h),\nonumber \\
\Lambda_{h} &=& \Lambda-\sqrt{m \eta}\arctan \frac{\Lambda}{\sqrt{m\eta}},
\label{HF}
\end{eqnarray} 
where $\Lambda_h$ is the dynamical length relevant to the renormalization transformation and depends on the many-body parameter $\eta$.
When $\eta=0$ as in vacuum, $g_2$ flows to the desired value of $4\pi a/m$.
With a finite $\eta$, we find that $g_2$ runs to the following value

\begin{eqnarray}
g_2^0 =\lim_{\Lambda_h \rightarrow 0} \frac{\tilde g_2(\Lambda_h)}{\Lambda_h}=\frac{4\pi a}{m}\frac{1}{1- \sqrt{2 m \eta} a}
\label{EI}
\end{eqnarray}
as $\Lambda_h$ becomes zero and all the $k\neq 0$ virtual states are included in the renormalization transformation.
The resultant $g^0_2$ is the effective interaction between condensed atoms after all non-condensed or virtual states are integrated out. 
Eq. (\ref{EI}) was proposed in Ref. \cite{Borzov11} as an effective interaction for condensed atoms. This is also fully consistent with the RSRT result
presented in section II.

At first sight, the structure of Eq. (\ref{EI}) appears to be very similar to 
the zero-density expression for the two-body running coupling constant $\tilde{g}_2(L)$ in Eq. (\ref{fsg2}). However, the physical implication is entirely 
surprising.
First of all, $g^0_2$, the effective interaction between condensed atoms now depends on $k_\eta=\sqrt{2m\eta}$; 
so it is now a function of $\Sigma-\mu$, or the density of the gases reflecting a many-body effect.
In the dilute limit,

\begin{eqnarray}
g_2^0=\frac{4\pi a}{m}(1+\sqrt{8\pi na^3}+...);
\end{eqnarray}
the first term stands for the Hartree-Fock energy and the second one yields the Lee-Huang-Yang type correction to the energy density of Bose gases.

And most importantly, unlike in vacuum where the zero energy effective interaction constant $4\pi a/m$ is always positive as far as $a$, 
the scattering length remains positive,
in condensates $g_2(\Lambda=0)$ is positive only in the dilute limit when $\sqrt{m\eta} a \sim \sqrt{na^3} \ll 1$.
When approaching the resonance, for a given $\Sigma-\mu$, the effective interaction between condensed atoms becomes negative before $a$ becomes infinity
as indicated in Fig. 4.
In other words, the presence of a condensate completely {\em alters} the flow of the coupling constant at the low energy limit; it changes the sign of 
the effective interaction constant near resonance.

The property of Bose gases near resonance is dictated by this change of the sign of interactions. In fact as a precursor of this, pure atomic condensates
lose metastability as seen in the following more elaborated discussion and diagrammatic resummation in section IV.
Microscopically, the change of sign of $g_2^0$ is correlated with and driven by the molecules' entering the condensate.
In the approximation employed here, the sign change occurs exactly when the molecules penetrate into the condensates at scattering length
$a_D$ (see Eq. (\ref{AD})). 

To further determine $\eta$ or $\mu$ and $\Sigma$ and understand the effect of the sign change of $g_2$ on the condensates, we should specify the RGEs with a boundary condition.
The following steps have to be carried out.
Once $g_2$ is found as a function of $\Sigma$ and $\mu$, one can apply it
to calculate $E(n_0,\mu)$, the energy density of the system with $n_0$ condensed atoms and non-condensed particles
at chemical potential $\mu$. Following the general thermodynamic relations, the chemical potential for the condensed particles $\mu_c$ should be

\begin{eqnarray}
\mu_c=\frac{\partial E(n_0,\mu)}{\partial n_0}, E=\frac{1}{2}g_2^0 n_0^2.
\label{muc}
\end{eqnarray}
And for the ground state we further require that the condensed atoms are in equilibrium with the non-condensed reservoir at chemical potential $\mu$.

\begin{equation}
\mu=\mu_c
\label{equili}
\end{equation}
which was first suggested by Pines and Hugenholtz \cite{Pines59}. 
One can verify that Eq. (\ref{EI}),(\ref{muc}) and (\ref{equili}) are identical to the corresponding self-consistent diagrammatic equations employed in Ref. \cite{Borzov11}.
More explicitly, one finds that for $g_2^0$,

\begin{eqnarray}
{g_2^0} n_0 +\frac{\partial g_2^0}{\partial \eta}\frac{\partial \Sigma}{\partial n_0}=\mu_c
\end{eqnarray}

\begin{figure}
\includegraphics[width=\columnwidth]{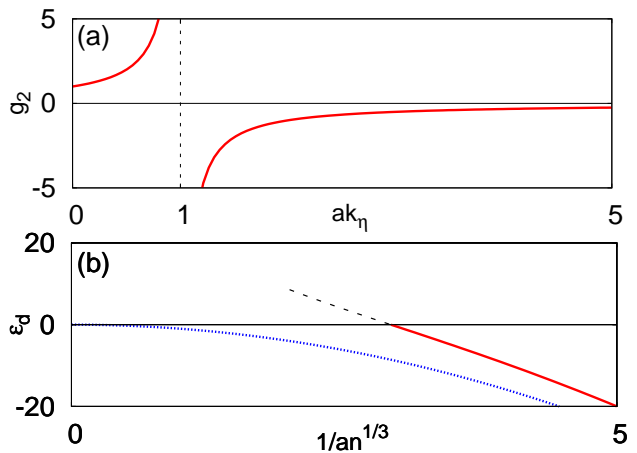}
\caption{(Color online) a) $g_2$ (in units of $4\pi a/m$) as a function of $k_\eta=\sqrt{2m\eta}$ at a given scattering length $a$; 
$\eta=\Sigma-\mu$ is a function of density and is equal to $4\pi na/m$ in the dilute limit.
Note that when $a k_\eta =\sqrt{8\pi na^3}\ll 1$ or in the dilute limit, $g_2$ approaches its vacuum value of $4\pi a/m$ 
but deviates from it substantially once $k_\eta a$ is
of order of unity. At resonance when $1/a=0$, $g_2$ is negative for any arbitrary $\eta$ implying attractive interactions between condensed atoms.
b) Illustration of the dimer energy (in units of $n^{2/3}/2m$) in the presence of a condensate (the upper curve).
The dashed line indicates dimers are no longer well defined because of the coupling to the continuum.
As a reference we also show the dimer energy in vacuum (the lower curve). Note in vacuum, the dimers reach zero energy right at resonance.}
\label{fig4}
\end{figure}

One can view Eq. (\ref{muc}) and (\ref{equili}) as a boundary condition for $\tilde{g}_2(\Lambda_h)$ in the RGE in Eq. (\ref{HF}) 
when $\Lambda_h=0$ and if $\partial \Sigma/\partial n_0$ is given.
To finally solve the equation self-consistently, one needs to supply Eq. (\ref{HP}) in the next section 
to further determine that $\partial \Sigma/\partial n_0=\eta/n_0$ and the set of equations
produced in this way are identical to the set in Ref. \cite{Borzov11} derived diagrammatically.

Here to illustrate the main features, we make a few further simplifications without losing the generality.
One is that we neglect the $n_0$-dependence in $\Sigma$ so that $\mu_c=g_2^0 n_0$.
Second is that we further approximate $n_0$ as $n$ because they are of the same order in the regime of our interest.
We then have a single parameter renormalization equation Eq. (\ref{HF}) with the following boundary condition

\begin{eqnarray}
g_2^0=\lim_{\Lambda_h \rightarrow  0} \frac{\tilde{g}_2(\Lambda_h)}{\Lambda_h}=\frac{\mu}{n}.
\label{BC}
\end{eqnarray}

Last, although $\eta=\Sigma-\mu$ in general should be $\beta \mu$ with $\beta$ being an unknown but smooth function of $a$, $n_0$ and $\mu$, in the dilute limit $\beta=1$.
Eq. (\ref{HP}) below implies that $\beta$ varies between $1$ in the dilute limit and $2/3$ in the fremionized limit that interests us.
So we can neglect its variation by simply setting $\beta=1$ for this part of discussion.
Eq.(\ref{HF}) and the boundary condition for the RGE in Eq.(\ref{BC}) now lead the following single parameter self-consistent equation for $\mu$,

\begin{equation}
\frac{\mu}{n}=\frac{4\pi a}{m(1- \sqrt{2m \mu} a)}
\label{SCE2}
\end{equation}
which, apart from a numerical prefactor in the denominator,
is identical to Eq. (\ref{SCE}) which was obtained empirically. 
The numerical solution of this is presented in Fig. 5.

\begin{figure}
\includegraphics[width=\columnwidth]{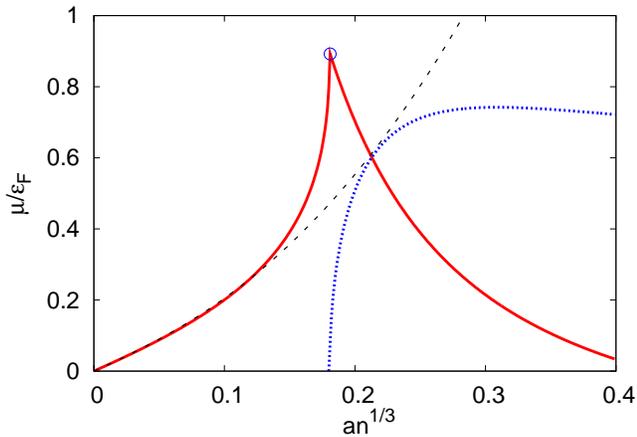}
\caption{(Color online) The numerical solution to the self-consistent equation Eq. (\ref{SCE2}).
The chemical potential (the real part) reaches the maximum (blue circle) when $n^{1/3}a=0.18$; beyond this point the chemical potential
develops an imaginary part (dotted line).
The dashed line is the chemical potential in the Lee-Huang-Yang theory.  
The smooth contributions from the three-body potential $g_3$ (not shown here) were studied in the previous diagrammatic calculations
\cite{Borzov11} and turn out to be around a few percent of the effect shown here. }
\label{fig5}
\end{figure}

Two essential features are shown in Fig. 5. First, the chemical potential reaches a maximum at $a_{cr}$
as a precursor of the sign-change of two-body interaction $g_2$ near resonance.
The value of maximum is around $89\% \epsilon_F$, 
very close to the values obtained in a constrained variational approach \cite{Song09} and in a diagrammatic resummation approach \cite{Borzov11}
(see Table I for details). Here $\epsilon_F=(6\pi^2)^{2/3}n^{2/3}/2m$ is the Fermi energy for a gas with the number density $n$.

And above the critical scattering length, the chemical potential develops an imaginary part.

\begin{eqnarray}
Im \mu=\frac{8}{(3\pi)^{2/3}}\epsilon_F (\frac{a}{a_{cr}}-1)^{1/2}
\end{eqnarray}
when the scattering length $a$ is increased slightly beyond the critical point
$a_{cr}=0.18 n^{-1/3}$.

A different renormalization group approach based in an atom-molecule model
was also applied in a previous study to understand
Bose gases near resonance~\cite{Lee10}.
Our results differ from theirs
in two aspects. First, in our approach, an onset instability
sets in near resonance even when the scattering length is positive, a feature that is absent in that previous study.
Second, when extrapolated to the limit of small $na^3$,
the results in Ref.~\cite{Lee10} imply a
correction of order of $\sqrt{na^3}$ to
the usual Hartree-Fock chemical potential but
with a negative sign, opposite to the sign of LHY corrections and/or our results.
The results of the self-consistent approach in Ref. \cite{Diederix11} are similar to the ones in Ref. \cite{Lee10} but differ from ours.
In table I, we make further comparisons by listing the main features in different approaches.

\begin{table}[ht]
\begin{tabular}{lcccc}
\hline\hline
    & Fermionized$^{*}$ & LHY & Efimov & Max. in $\mu$, \\
   &  & effect  & physics  & instability \\
\hline
Cowell {\it et al.}, & Yes & No & No & No \\
2002\cite{Cowell02} & (2.92$\epsilon_F$)$^{\dagger}$ & & & \\
\hline
Song {\it et al.}, 
& Yes & No & No & No \\
2009\cite{Song09} & ($0.80\epsilon_F$) & & & \\
\hline
Lee {\it et al.},
& Yes & No$^{\dagger\dagger}$ & No & No\\
2010\cite{Lee10} & ($0.66\epsilon_F$) & & & \\
\hline
Diederix {\it et al.}, 
& Yes & No$^{\dagger\dagger}$ & No & No\\
2011\cite{Diederix11} & ($0.83 \epsilon_F$) & & & \\
\hline
Borzov {\it et al.},
& Yes & 
Yes & Yes & Yes$^{\dagger\dagger\dagger}$ \\
2012\cite{Borzov11} & ($0.93\epsilon_F$) & & & \\
\hline \hline
\end{tabular}
\caption{Comparison of Different Theory Approaches \\ \\
* The lower bound of $\mu$ was measured to be around $0.44 \epsilon_F$ in the ENS experiment \cite{Navon11}.\\
$\dagger$
The value in the bracket indicates the estimated chemical potential; same below.
The estimated chemical potential $2.92\epsilon_F$ exceeds the result for a completely fermionized gas.
\\
$\dagger\dagger$
In the field theory approaches there,
the signs of the correction of order of $\sqrt{na^3}$ are opposite to the LHY effect. 
However, the LHY effect was reproduced in the numerical program in Ref.\cite{Diederix11}.\\
$\dagger\dagger\dagger$ 
This is seen both in the diagrammatic resummation and the RG approach outlined here.
Note that $0.93 \epsilon_F$ is for a range of three-body parameters relevant to cold atoms.}
\end{table}

\section{Diagrammatic resummation: A self-consistent approach}

From a phenomenological point of view, it is quite appealing to generalize the
self-consistent coarse grain relation Eq. (\ref{SC}) by further taking into account the three-body effective interaction $g_3$;

\begin{eqnarray}
\mu =n g_2(\xi, a; n)+\frac{n^2}{2} g_3(\xi,a; n)
\end{eqnarray}
where $g_{2,3}(L,a;n)$ are the renormalized two- and three-body interaction constants respectively at length scale $L$ 
and $\xi^{-1}=\sqrt{2m\mu}$.
If one can calculate these renormalized quantities, then one is able to obtain $\mu$ which includes the effect of $g_3$.
We have proceeded further from here using the renormalization group equations similar to what was discussed in section II, III;
they yield qualitatively the same results as the diagarmmatic approach below. However, when benchmarking against the dilute gas theory, the diagrammatics turn out to be numerically superior;
the diagrammatic resummation used in our previous paper reproduces $99.96\%$ 
of Lee-Huang-Yang corrections in the dilute limit. For this reason, we are going to
outline the framework of the diagrammatic calculations and briefly comment on the results.

The Hamiltonian for a condensate with a short range interaction is

\begin{eqnarray}
H&=&\sum_{\bf k} (\epsilon_{\bf k} -\mu) b_{\bf k}^\dagger b_{\bf  k}
+ 2 U_0 n_0 \sum_{\bf k} b^\dagger_{\bf  k} b_{\bf  k}
\nonumber \\
&+&\frac{1}{2} U_0 n_0\sum_{\bf k}
b^\dagger_{\bf  k}
b^\dagger_{-\bf  k}+
\frac{1}{2}U_0 n_0 \sum_{\bf k} b_{\bf  k}b_{-\bf  k}
\nonumber \\
&+&\frac{U_0}{2\sqrt{\Omega}}\sqrt{n_0}
\sum_{{\bf k'},{\bf q}} b^\dagger_{\bf q} b_{\bf  k'+\frac{\bf q}{2}}
b_{-\bf  k'+\frac{\bf q}{2}}+h.c.
\nonumber \\
&+&\frac{U_0}{2\Omega} \sum_{{\bf k}, {\bf k'},{\bf q}} b^\dagger_{\bf k+\frac{\bf q}{2}} b^\dagger_{-\bf  k+\frac{\bf q}{2}} b_{\bf  k'+\frac{\bf q}{2}}
b_{-\bf  k'+\frac{\bf q}{2}}+h.c.
\end{eqnarray}
Here $\epsilon_{\bf k}={\bf k}^2/2m$ and
the sum is over non-zero momentum states.
$U_0$ is the strength of the contact interaction related to the scattering length
$a$ via $U_0^{-1}=m (4\pi a)^{-1} -\Omega^{-1} \sum_{\bf k} (2\epsilon_{\bf k})^{-1}$, $\Omega$ is the volume. 
$n_0$ is the number density of the condensed atoms and $\mu$ is the chemical potential, both of which are functions of $a$ and to be determined self-consistently.

In the diagrammatic approach, we first define the chemical potential of non-condensed particles as $\mu$ and the number density of condensed atoms is $n_0$.
The energy density can be calculated as $E(n_0,\mu)$ (see below).
Then one should have the following set of self-consistent equations for a gas with total number density $n$,

\begin{eqnarray}
\mu_c &=& \frac{\partial{E(n_0,\mu)}}{\partial n_0} \nonumber \\
n &=& n_0 -\frac{\partial{E(n_0,\mu)}}{\partial \mu} \nonumber \\
\mu &=&\mu_c
\label{SCD}
\end{eqnarray}
where $\mu_c$ is the chemical potential for the condensed atoms and has to be equal to the chemical potential $\mu$ in equilibrium.
We further introduce the self-energy $\Sigma$ of non-condensed atoms or virtual particles to facilitate the calculation of $E(n_0,\mu)$ that now explicitly depends on
$\Sigma(n_0,\mu)$. Thus, Eq. (\ref{SCD}) has to be further
supplemented by

\begin{eqnarray}
\Sigma(n_0,\mu)=\mu_c(n_0,\mu) +\frac{\partial \mu_c}{\partial \ln n_0}
\label{HP} 
\end{eqnarray}
which can be proven in the same fashion as the Pines-Hugenholtz theorem \cite{Pines59}.
Eq. (\ref{SCD}) and (\ref{HP}) have been applied to obtain the chemical potential in 3D Bose gases near resonance \cite{Borzov11}.

Calculations of $E(n_0,\mu)$ for a given $\Sigma(=\eta+\mu)$ can be carried out diagrammatically.
If we restrict ourselves to the virtual processes involving only two or three excited atoms and truncate the Hilbert space accordingly, then 
diagrammatically we only need to collect the diagrams which contribute to the effective two- and three-body interaction constants $g_{2,3}$.
As far as the chemical potential is concerned, this truncation turns out to be highly precise in the dilute limit. The result is listed below.
The mass $m$ is set to be one.

\begin{eqnarray}
E(n_0,\mu)& =&\frac{1}{2}n_0^2 g_2(2\eta) +\frac{1}{3!}n_0^3 g_3(3\eta) \nonumber \\
g_2(2\eta)&=&{4\pi a} \frac{1}{1-\frac{2}{\pi}\sqrt{2\eta} a} \nonumber \\ 
g_{3}(3\eta)&=&6 g^2_2(3\eta) Re \frac{2}{\pi}\int dq\frac{K(-2\eta;0,q)q^{2}}{\sqrt{\frac{3}{4}q^{2}+3\eta}-\frac{1}{a}} G^{'}_{3}(-3\eta,q)
\nonumber \\
\label{gs}
\end{eqnarray}
where
$G^{'}_{3}(-3\eta, p)$ is a solution of the following integral equation

\begin{widetext}
\begin{eqnarray}
G^{'}_{3}(-3\eta,p)=\frac{2}{\pi}\int dq\frac{K(-3\eta;p,q)q^{2}}{\sqrt{\frac{3}{4}q^{2}+3\eta}-\frac{1}{a}}
[\frac{-1}{q^{2}+2\eta}+ G^{'}_{3}(-3\eta,q)].
\end{eqnarray}
\end{widetext}
And $-1/2K(-3\eta; p, q)$ is again the one-particle Green's function projected to the $S$-wave channel; it is defined as

\begin{eqnarray}
K(-3\eta; p,q)=\frac{1}{pq}\ln \frac{p^2+q^2+3\eta +pq}{p^2+q^2+3\eta-pq}.
\end{eqnarray}
In the appendix, we outline the major steps in the loop summation which leads to $E(n_0,\mu)$.

The numerical solution to these self-consistent equations was shown in Ref. \cite{Borzov11}
and they are qualitatively the same as the solution to
the self-consistent RGE for $g_2$ and we are not going to repeat here \cite{note}.
Here we want to make a few further comments on the resummation technique.

First, $g_2$ defined this way is an effective two-body interaction renormalized by condensates and includes 
a subset of $N$-body interactions defined in the vacuum.
At one-loop level, it yields the most dominating contribution; the residue effects are from
the irreducible $N=4,6,...$-body interactions which contains less than one thousandth of the total contribution.

Second, the three-body contribution in our self-consistent approach appears to be around a few percent
and numerically insignificant. Since when compared to $g_2$, the contribution from $g_3$ in the dilute limit as well as near resonance is small as shown previously \cite{Borzov11}, 
it is reasonable to conjecture that further inclusion $g_{4,5,...}$ wouldn't change our result presented here
in a substantial way. The truncation of the energy density expression at $g_3$ should be accurate enough
for all the practical purposes of studying Bose gases near resonance. 
We hope these statements can be tested 
in precision measurements of chemical potentials
as well as in future quantum Monte Carlo simulations.

Third, the energy density expression in Eq. (\ref{gs}) becomes exact in the limit where only the processes involving two or three virtual atoms are allowed.
Effectively, this is equivalent to truncating the Hilbert space and including the correlations up to the trimer channel.

\section{Conclusions} 
The RGE approach is instrumental to our understanding of the emergent phenomena in quantum few- and many-body systems. The application to Bose gases near resonance
perhaps is another example of what a simple RGE transformation can lead to.
We have applied this approach to understand the nature of Bose gases
near resonance and
found that energetically, the Bose gases close to unitarity are nearly {\it fermionized} before an  onset instability sets in,
{\it i.e.} the chemical potentials of the Bose gases
approach that of the Fermi energy of a Fermi gas with equal mass and density.
Beyond the instability point, the chemical potential has an imaginary part indicating strong hybridization with molecules.

The model we have employed to study the Bose gases near resonance is a short range {\em attractive} potential which has a range much shorter than the interatomic distance
of the gases
or effectively a contact potential. This is a very good approximation of real physical interactions between cold atoms.  
If the potential is a short range but repulsive one, then Bose gases are always in the dilute limit because the scattering 
lengths are bounded by the range of interactions, disregarding the strength of potential. For bosons interacting with a repulsive potential but with a range comparable to
the interparticle distance, we should anticipate the physics in this limit to be very similar to what happens in liquid $^4He$
\cite{London54,Landau41,Feynman54,Penrose56,Affleck11,Nozieres90}. 
The excitation spectrum should develop
roton minima that imply strong short range crystal correlations. When the range of interactions is further increased, 
eventually there should be a quantum transition to a crystal where
all bosons are depleted from the condensate. The physics of repulsive bosons and liquid $^4He$ belong to a different universality class which fundamentally differs from
what we described in this article, i.e. the properties of nearly fermionized Bose gases near resonances with a contact interaction.

We want to thank Ian Affleck, Jean-Sebastien Bernier, Dmitry Borzov, Junliang Song, Joseph Thywissen, Shizhong Zhang for many inspiring discussions and Chen-Ning Yang for 
warm encouragement.
The points of view in this article are based on a series of seminars given by one of the authors (F.Z.) at Harvard University, INT, University of Washington,
University of Hamburg, IASTU, Tsinghua University, Zhejiang University and Fudan University during his sabbatical leave in 2011;
F.Z. wants to thank Eugene Demler, Klaus Sengstock, Hwa-tung Nieh, Qijing Chen and Yu Shi for arranging his visit and their hospitalities. 
The exciting discussions in 2011 have shaped our current understanding of the Bose gases near resonance.
Finally, F.Z. wants to thank Zheng-Yu Weng for numerous wonderful walks in Tsing-Huwa-Yuan, and Qijing Chen for a very pleasant hike in
the Botanical Garden by the West Lake, Hangzhou.  
This work is in part supported by NSERC (Canada), Canadian Institute for Advanced Research, and Izaak Wlaton Killam Memorial Fund for Advanced Studies.

\begin{widetext}

\section{Appendix: Two- and Three-Body Scattering Amplitudes in a condensate and $g_{2,3}$}

In the following, we show the explicit calculation of the three-atom scattering amplitude $G_3$, and
$E(n_0,\mu)$ for a given $\Sigma(=\eta+\mu)$ by adding the diagrams with minimum number of virtual particles involved. 
The model Hamiltonian could be written as:

\begin{eqnarray}
H&=&\sum_{\bf k} (\epsilon_{\bf k} -\mu) b_{\bf k}^\dagger b_{\bf k}
+ 2 U_0 n_0 \sum_{\bf k} b^\dagger_{\bf k} b_{\bf k}
+\frac{1}{2} U_0 n_0\sum_{\bf k}
b^\dagger_{\bf k}
b^\dagger_{-\bf k}+
\frac{1}{2}U_0 n_0 \sum_{\bf k} b_{\bf k}b_{-\bf k}
\nonumber \\
&+&\frac{U_0}{2\sqrt{\Omega}}\sqrt{n_0}
\sum_{{\bf k'},{\bf q}} b^\dagger_{\bf q} b_{\bf k'+\frac{\bf q}{2}}
b_{-\bf k'+\frac{\bf q}{2}}+h.c.
+\frac{U_0}{2\Omega} \sum_{{\bf k}, {\bf k'},{\bf q}} b^\dagger_{\bf k+\frac{\bf q}{2}} b^\dagger_{-\bf k+\frac{\bf q}{2}} b_{\bf k'+\frac{\bf q}{2}}
b_{-\bf k'+\frac{\bf q}{2}}+h.c., \label{H}
\end{eqnarray}
where the sum is over non-zero momentum states. $U_0$ is the strength of the contact interaction which is related to scattering length $a$ as:

\begin{equation}
{1\over U_0}={m\over 4\pi a}-{1\over \Omega} \sum_{k} {1\over 2\epsilon_k},
\end{equation}
where $\Omega$ is the volume. Taking into account only two and three body interactions, the energy density could be written as:

\begin{equation}
E(n_0,\mu)={1\over 2}n_0^2g_2(2\eta)+{1\over 3!}n_0^3g_3(3\eta),
\end{equation} 
where $g_2$ and $g_3$ are irreducible 2 and 3-body potentials respectively. $g_2$ could be found by writing the Bethe-Salpeter equation as:

\begin{equation}
g_2(2\eta)^{-1}=U_0^{-1}-i\int {d\omega\over 2\pi}{d^3 {\bf k} \over (2 \pi)^3}G(\omega,{\bf k})G(-\omega,-{\bf k}),
\end{equation}
where $G(\omega,{\bf k})^{-1}=\omega-\epsilon_{\bf k}-\eta+i\delta^+$ is the interacting Green's function. So, the two-body potential could be obtained as:

\begin{equation}
g_2(2\eta)={4\pi \over {1\over a}-\sqrt{2m\eta}}\label{g2}
\end{equation}

Similarly, $g_3$ could be estimated by summing up all N-loop diagrams with 3 incoming and outgoing lines which are depicted in Fig. 3(b). 
We consider a general case where three incoming momenta are ${\bf k}_1={\bf p}/2-{\bf q}$, ${\bf k}_2={\bf p}/2+{\bf q}$ and 
${\bf k}_3=-{\bf p}$, and outgoing ones are ${\bf k}'_1={\bf p}'/2-{\bf q}'$, ${\bf k}'_2={\bf p}'/2+{\bf q}'$ and ${\bf k}'_3=-{\bf p}'$. The scattering amplitude between theses states is then given by $A_3(E;{\bf p},{\bf p}')$. At the tree level, the effective 3-particle interaction is:

\begin{equation}
\Gamma^{(0)}={f_0\over E-\omega_{in}-\omega_{out}-\epsilon_{{\bf p}+{\bf p}'}-\eta+i\delta^+}, \label{gamma0}
\end{equation}
where $\omega_{in}$ and $\omega_{out}$ are frequencies of lines with momentum ${\bf k}_3$ and ${\bf k}'_3$ respectively. Furthermore, $f_0$ is the product of the perturbation factor $f_p$, vertex factor $f_v$ and symmetry factor $f_s$ which will be explained later. 

To keep the notations simple, we set $m=1$ from now on.
We consider on-shell limit and substitute $\omega_{in}$ and $\omega_{out}$ by $p^2/2-\eta$ and $p'^2/2-\eta$ respectively. To project into the $S$-wave channel, we take the average over all directions,

\begin{equation}
\overline{\Gamma}^{(0)}={-f_0\over 2pp'} ln\Bigg({p^2+p'^2+pp'+3\eta-E\over p^2+p'^2-pp'+3\eta-E}\Bigg)\equiv {-f_0\over 2} K(E-3\eta;p,p'),
\end{equation}
where we have defined the kernel $K$ as:

\begin{equation}
K(E-3\eta;p,p')={1\over pp'}ln\Bigg({p^2+p'^2+pp'+3\eta-E\over p^2+p'^2-pp'+3\eta-E}\Bigg)
\end{equation}

{\it Perturbation factor}, $f_p$, comes from the expansion of the exponential of interaction term in Hamiltonian in perturbation method. The diagrams with $l$ vertices could be written as:

\begin{equation}
{1\over l!}(V_A+V_B)^l ,
\end{equation}
where $V_A$ and $V_B$ stand for interaction terms corresponding to different types of vertices namely $A$ and $B$. 
For example, for a diagram with 2 vertices of type $A$ and one vertex of type $B$, $f_p$ is the factor in front of $V_A^2V_B$ term in numerator of the above equation divided by $l!$. In the vacuum case, where all the vertices are the same and all the lines can have non-zero momenta, $f_p$ is simply equal to $(1/l!)$.

{\it Vertex factor}, $f_v$, is defined as the product of the factors in front of $g_2$ for different vertices shown in Hamiltonian. For vacuum case, all the vertices have $1/2$ factor and $f_v$ is equal to $(1/2)^l$.

The last factor is {\it symmetry factor}, $f_s$ which shows the number of identical diagrams generated for a given number of vertices. For vacuum case, $f_s=l!\times 4^l$ where $l!$ 
shows the number of permutations of vertices and $4^l$ is the number of different ways of connecting vertices (2 for incoming lines and 2 for outgoing lines). 

So, in general the prefactor appearing in $\Gamma^{(n)}$ where $n$ is the number of loops ($n=l-2$) would be

\begin{equation}
f_n=2^{(n+2)};
\end{equation}
and $f_0=4$.

$\Gamma^{(1)}$ could be written in terms of kernel defined above as:

\begin{equation}
\Gamma^{(1)}=8 \int {d^3k\over (2\pi)^3} (-{1\over 2} K(E-3\eta;p,k))g_2(E-3\eta-{k^2\over 2};k)(-{1\over 2} K(E-3\eta;k,p')),\label{gamma1}
\end{equation}
where the integral over internal frequency has taken. $g_2(\omega;Q)$ has the following form in 3D:

\begin{equation}
g_2(\omega;Q)={4\pi \over {1\over a}-\sqrt{{Q^2\over 4}-\omega}},
\end{equation}
where above equation reduces to Eq. (\ref{g2}) in the limit of zero energy and momentum. The effective three-body interaction then could be obtained by summing over $\Gamma^{(n)}$s:

\begin{eqnarray}
\Gamma_{eff}&=&-2K(E-3\eta;p,p')+{4\over \pi}\int dk k^2 {K(E-3\eta;p,k)\over {1\over a}-\sqrt{{3k^2\over 4}-E+3\eta}}\times K(E-3\eta;k,p')\nonumber\\
&-&{8\over \pi^2}\int dk k^2 \int dk' k'^2  {K(E-3\eta;p,k)\over {1\over a}-\sqrt{{3k^2\over 4}-E+3\eta}}\times {K(E-3\eta;k,k')\over{1\over a}-\sqrt{{3k'^2\over 4}-E+3\eta}}\times K(E-3\eta;k',p')+\dots \label{gammaeff}
\end{eqnarray} 

The sum of above infinite series leads to the following integral equation for scattering amplitude $A_3$:

\begin{equation}
A_3(E;p,p')=-2K(E-3\eta;p,p')-{2\over \pi}\int dk k^2 {K(E-3\eta;p,k)\over {1\over a}-\sqrt{{3k^2\over 4}-E+3\eta}} A_3(E;k,p')\label{integral}
\end{equation}
One then obtains the reduced scattering amplitude   
$G_3(E; {\bf p})=A_3(E;{\bf p},0)$ in Eq. (\ref{G3IE}). 

For the calculation of $g_3$ for condensates, one has to exclude the tree level diagram that no longer exists because of momentum conservation.
The sum of the rest infinite series leads to the following integral equation for the scattering amplitude 
$A_3(E;{p},{p}')$,

\begin{equation}
A_3(E;{p},{p}')={2\over \pi}\int dk {K(E-3\eta;p,k)k^2\over {1\over a}-\sqrt{{3k^2\over 4}-E+3\eta}}\Bigg(\frac{1}{2} K(E-3\eta;k,p')-A_3(E;{k},{p}')\Bigg)
\label{integral2}
\end{equation}

The above scattering amplitude could then be applied to calculate the scatterings between condensed atoms
when setting $E$, ${\bf p}$ and ${\bf p}'$ to be zero in the above equation but 
with two further modifications. 
The first change is the numerical factor in front of the effective interactions. This factor in condensate case is $1/4$ of 
the factor in vacuum case, 
because there is a $2\times 2$ factor for changing the external legs of the external vertices for non-zero incoming momenta.
So we will get the same integral equation, but the first term in the bracket of the integrand of Eq. (\ref{integral2}) would be substituted with $1/2 K$.

The second change would be in the shift of the energy. If we set the momentum of external legs to zero from the beginning, In the on-shell limit there is no shift of 
the energy for $\omega_{in}$ and $\omega_{out}$ in Eq. (\ref{gamma0}). 
So, the energy of the first and the last kernel in all the terms of the Eq. (\ref{gammaeff}) other than the tree level term would be $E-2\eta$ in condensate case. 
Subtracting the one-loop contribution which has already been counted in
the renormalized $g_2$ and taking into account the above two modifications, we obtain $g_3$ via the relation shown in Eq. (39), (40)~\cite{note}.
  
\end{widetext}


\begin{thebibliography}{References}
\bibitem{note}There are two ways of estimating $G_3$ which slightly differ from each other; the difference is due to
singular behavior of the Green's function at $k=0$ in the Hatree-Fock approximation as commented on in Appendix C of Ref.\cite{Borzov11}. Here we evaluate $G_3$ by first setting all
external lines to be the condensed atoms.
\end{thebibliography}
\end{document}